# Modern Data Analytics Approach to Predict Creep of High-Temperature Alloys

D. Shin[1]*, Y. Yamamoto[1], M. P. Brady[1], S. Lee[2], and J. A. Haynes[1]

*[1]Materials Science and Technology Division, Oak Ridge National Laboratory, Oak Ridge, TN 37831*

*[2]Computer Science and Mathematics Division, Oak Ridge National Laboratory, Oak Ridge, TN 37831*

## Abstract

A breakthrough in alloy design often requires comprehensive understanding in complex multi-component/multi-phase systems to generate novel material hypotheses. We introduce a modern data analytics workflow that leverages high-quality experimental data augmented with advanced features obtained from high-fidelity models. Herein, we use an example of a consistently-measured creep dataset of developmental high-temperature alloy combined with scientific alloy features populated from a high-throughput computational thermodynamic approach. Extensive correlation analyses provide ranking insights for most impactful alloy features for creep resistance, evaluated from a large set of candidate features suggested by domain experts. We also show that we can accurately train machine learning models by integrating high-ranking features obtained from correlation analyses. The demonstrated approach can be extended beyond incorporating thermodynamic features, with input from domain experts used to compile lists of features from other alloy physics, such as diffusion kinetics and microstructure evolution.

---

* Corresponding Author
Email address: shind@ornl.gov (Dongwon Shin)

Keywords: high-temperature alloys, creep, correlation analysis, machine learning, features, computational thermodynamics

## 1. Introduction

The design of advanced alloys for high-temperature applications requires comprehensive understanding in multi-component (e.g., Fe, Ni, Cr, Mn, Si, Al, Co, Mo, Nb, Ti, V, W, B, C, N, etc.), multi-phase (e.g., austenite, ferrite, intermetallics, carbides, nitrides, etc) and multi-physics (e.g., thermodynamic, kinetics, and mechanics) as well as concurrent consideration of multiple descriptors at different length scales (i.e., atomic, nano-, micro-, meso- and macroscale). However, it is daunting, cost-prohibitive and time-consuming to promote such complex understanding solely from a series of experimental investigation as the search space is extremely high-dimensional. Hence, the high-temperature alloy design community has been constantly developing theoretical capabilities to accurately and rapidly predict alloy properties to narrow down the search space.

Models used to predict materials properties can be roughly categorized into physics-based and analytical models. Physics-based models (e.g., density functional theory and phase-field simulations) can promote understanding at given length scales (i.e., atomistic and microstructure). With the recent remarkable progress in supercomputing, the speed and scale of simulations of physics-based models have been greatly enhanced [1]. However, they are often limited to low-order model systems due to the finite computing power even with modern resources or lack of input parameters to represent realistic higher-order systems. Although these physics-based models can greatly facilitate understanding of underlying mechanisms, they are not yet mature enough to suggest prototypical chemistry of complex, multi-component/multi-phase alloys.

On the other hand, analytical surrogate models are better suited for alloy design as they can better handle multi-component systems with much less computing power [2–6]. There are a number of legacy (mainly based on neural networks) [7–9] and recent efforts (based on machine learning) [10–13] to apply data analytics approaches for developing predictive capabilities toward multi-component/multi-phase alloy design. These approaches usually require a large volume of experimental data to ensure the fidelity of trained models. In this regard, accumulated alloy property data over decades can potentially be leveraged.

However, these analytics-based approaches are built on the assumption that the microstructures/chemistries and the underlying mechanisms that control alloy properties remain the same, thus it is risky to extrapolate outside the range of the data which trained the algorithm. Further, determining the level of confidence in data compatibility can be a significant challenge, particularly when data for a specific type of alloy are obtained from multiple sources, often with lack of adequate specificity in alloy composition, microstructure and processing history, as well as potential variations in approach, consistency and accuracy in measuring and reporting properties data. In addition, previous efforts have used only simple and superficial descriptors (e.g., nominal bulk alloy elemental compositions, with limited to no alloy processing condition history descriptors) as input features. Hence, such surrogate models are often referred to as 'black box' as they provide little if any insight into developing a fundamental hypothesis that can be transferred to other classes of alloys or operating conditions. These simple feature-based data analytics approaches add value, however, they are primarily constrained to only incremental improvements in alloy design. Thus, it is necessary to develop advanced data analytics approaches that more systematically incorporate physically/chemically meaningful scientific features to facilitate potential alloy hypotheses generation in the training of surrogate models.

Herein, we demonstrate a theoretical workflow that combines consistently-obtained experimental data and high-throughput interrogation of multi-component/multi-phase materials models to populate scientific features within the context of modern data analytics. We use an example of creep properties of a developmental, high-temperature class of Fe-base, alumina-forming austenitic (AFA) stainless-steel alloys consistently measured over the past decade [14–20]. Analyzed (not nominal) bulk alloy elemental compositions and heat treatment/creep evaluation temperatures of AFA alloys have been used as input for calculating alloy features with a computational thermodynamic approach that are closely related to underlying strengthening mechanisms. We then perform extensive correlation analyses to identify features that significantly influence creep behavior of AFA alloys. We scrutinize the rankings of simple/superficial and synthetic/scientific features obtained from correlation analyses on the training performance of various machine learning models. Finally, we discuss the importance of input from domain experts (high-temperature alloy designers in this example) to identify key feature types to be included in the correlation analysis and machine learning. We also demonstrate interrogating the developed surrogate models to recognize the most significant trends and achieve new insights to permit more rapid and cost-effective development of alloys with improved capabilities. The proven approach here can also be extended to other types of scientific features, for example diffusion kinetics, and coarsening related behaviors which are also well established to play an important role in creep resistance.

## 2. Experimental Data and Computational Methods

### 2.1. Experimental Alloy Data

Creep-resistant, AFA stainless-steel alloys are a new, developmental class of Fe-base heat-resistant structural alloys, initially targeted for use in the ~500-900°C temperature range [14–20]. They possess a combination of superior oxidation resistance through protective alumina-scale formation rather than conventionally used chromia scales, and high creep strength via controlled second-phase precipitation strengthening at elevated temperatures. Development of AFA alloys was pursued in the past decade [15–17,21] with an initial target for use in heat-exchanger components in energy conversion and combustion system applications [18]. The design strategy of AFA alloys was based on (1) properly balanced additions of Fe, Cr, Al and Ni to maintain a single austenite-phase matrix for high temperature creep strength; (2) balance of Al, Cr, Ni, and Nb additions to yield protective alumina-scale formation; (3) use of minor alloying additions (such as Mn, Si, Nb, Mo, W, Ti, Zr, Hf, Y, Cu, C, and B) to improve stability of the austenite-phase matrix, increase the effect of solution hardening, maximize strengthening second-phases (carbides, intermetallic-phases), while also minimizing any negative impacts on formation of the protective alumina scale [15,17,19].

AFA alloys have been designed to use MC, $M_{23}C_6$, Laves and/or $L1_2$ strengthening precipitates in an austenite single-phase matrix. Over the past decade more than 100 lab-scale arc-cast (0.1 to 0.5 kg) and pilot scale industrial vacuum cast (15 kg, with several compositions at 200 kg and 4000 kg) AFA alloys were manufactured and evaluated for creep (and oxidation) resistance, with nominal composition range of Fe-(12-32)Ni-(12-20)Cr-(0-5)Al-(0-3.3)Nb-(0-1)Ti-(0-1)V-(0-2)Mo-(0-2)W-(0-1)Si-(0-12)Mn-(0-3)Cu-(0-0.5)Zr-(0-0.2)Y-(0-0.1)B, in wt% [14–20]. Bulk alloy compositions were analyzed by a combination of inductively-coupled plasma (ICP), inert gas fusion analysis and combustion techniques. The analyzed alloy compositions were used in the computational thermodynamic calculations used to populate scientific features into the AFA dataset. Figure 1 represents typical microstructures of AFA alloys after aging (Figure 1a) or creep-rupture testing (Figure 1b) at 750°C [19] which consist of multiple second-phase particles including nano-scale intermetallic compounds and carbides formed on the grain boundary and inside the FCC-Fe matrix. They are key factors to improve the creep performance of AFA alloys, and their individual phase compositions, volume fractions, and degree of supersaturation from solutionizing condition to precipitation at service temperatures are considered to be the important scientific features in the current study.

Quantified microstructural information such as strengthening phase size, distribution, and preferential locations, both at grain boundaries and intra grain regions, were not consistently available across the developmental AFA alloy experimental dataset to include as scientific feature inputs. Such information is clearly desirable for understanding and predicting creep behavior, but are not experimentally available in many alloy datasets. Theoretical prediction of high-quality, quantified microstructural evolution is also not yet readily available, especially in developmental alloy composition ranges. The high-throughput thermodynamic phase equilibria calculations used in the present work (as described in detail in the following section), including aspects of supersaturation (which may influence precipitate size, for example), are a good first step beyond simple features, which can only indirectly be linked to alloy properties such as creep behavior.

[Figure 1 about here]

The AFA alloys were hot/cold worked to sheet/plate product using similar processing parameters, with creep (and oxidation) evaluation usually conducted in the as-solution treated condition, typically 1200°C, although ranging from 1100-1250°C depending on specific alloy melting point and strengthening design strategy. AFA alloy creep test results that utilized cold work to enhance strengthening precipitate formation [14,15,19,20] were specifically excluded from the AFA dataset used for data analytics (see Figure 2). Creep test sample configurations included both sub-sized and standard sized configurations according to ASTM E8. Although foil form AFA creep tests were also conducted, that data was excluded from the AFA dataset, as foils often exhibited deviation from sheet/plate form creep behavior due to their more extensive processing.

High-load, short duration creep test conditions were typically used to permit rapid feedback to the AFA alloy design effort: usually 650°C/250MPa, 700°C/170MPa, 750°C/100MPa, 800°C/80MPa, although select additional conditions were also occasionally used (see Figure 2). Creep rupture lifetimes were in the range of ~20 to 7000 h, with most falling in the ~100-1500 h range. We use Larson-Miller Parameters (LMP) to represent the creep behavior of AFA alloys:

$$LMP = (T \ [°C]+273) \times (C + \log t_{rupture} \ [h]), \ C=20 \qquad (1)$$

It is important to note that all alloys were not tested for all conditions. Creep testing was pragmatically dictated by alloy performance, and the need for rapid feedback for alloy design efforts. Property datasets have typically not been created with the use of data analytics in mind, and complete history and knowledge of the source property data is critical to achieving a high-quality, consistent dataset for analysis. The list of elements, temperature, stress and phases of the AFA alloys' experimental creep dataset and computational thermodynamic calculation are summarized in Table 1.

[Table 1 about here]

[Figure 2 about here]

2.2. Synthetic/scientific alloy features via high-throughput computational thermodynamics

We wish to go beyond previously attempted correlating bulk alloy composition and creep test features (stress and temperature) with LMP of AFA alloys within the context of data analytics. For example, individual compositions of Nb and C additions for a given AFA alloy can potentially be associated with corresponding LMP values from the creep test in alloys primarily utilizing NbC strengthening precipitates. However, bulk alloy composition does not directly control creep, rather analyzed bulk alloy compositions can be used to predict phase equilibria and, to some degree, microstructure, which more directly control creep behavior. The more relevant features rather than simple elemental bulk alloy compositions would be the phase/volume fraction and distribution of the precipitates at pertinent temperatures, grain/particle size, location of grain boundary. It is possible to experimentally determine such critical microstructure-related features that are highly relevant to mechanical properties of alloys, but it is cost-prohibitive and time-consuming to collect such information of every sample. Hence, we have used a computational thermodynamics approach [22] in a high-throughput manner to rapidly generate a subset of key phase equilibria/microstructural alloy features, which could potentially be better correlated with experimentally measured creep data, represented as LMPs. We have used the state-of-the-art TCFE9 database as implemented in the Thermo-Calc software package to compute relevant alloy features from analyzed alloy compositions. Populated synthetic alloy features are summarized in Table 2.

[Table 2 about here]

The scientific alloy features for calculation were selected in consultation with the AFA alloy design experts. The AFA alloys were originally designed primarily based on bulk alloy composition trends associated with $M_{23}C_6$, MC, Laves, and $L1_2$ strengthening phases, with an emphasis on phase/volume fractions at the intended use creep test temperature, and degree of supersaturation from solutionizing temperature to use temperature. These phases, as well as all other phases either identified in alloy characterization efforts or predicted using the TCFE9 database were included (e.g. FCC matrix, B2, sigma, metal borides, etc. See Table 1). In addition to superficial bulk alloy composition (in at%), solutionizing temperature (T1), creep test temperature (T2), dT (solutionizing minus creep test temperature), features related to phase/volume fraction of every predicted phase at T1 and T2, the degree of supersaturation from T1 to T2 for every phase, activity and composition (at%) of every element in every phase were also calculated. Two additional 'engineered' features for Nb:C ratio (at %) and phase fraction times element concentration for every phase at both T1 and T2 were also provided (see Table 2). Obviously, inclusion of these hundreds of advanced features is beyond the capability of human consideration without the aid of advanced data analytics. However, even the present approach is limited by the fact that a detailed microstructural characterization comparison between predicted and actual phase features, fractions, and compositions was not feasible within this study. The AFA alloys are a challenge for commercial thermodynamic databases to predict accurately as they fall near the interface between Fe- and Ni- base alloys and have phases such as B2 and $L1_2$ in conjunctions with metal carbides and borides that are not typically encountered in stainless steels due to their relatively high Al content (2.5 to 4 wt.%). That said, the observed trends from TCFE9 database in general were usually qualitatively reasonable with the available characterization findings for selected alloys conducted in previous work.

2.3. Correlation analysis between considered features and AFA LMPs

The AFA dataset used in the current work is high-dimensional: 82,502 data points consist of 466 input features/1 target property (LMP) and 166 alloy instances. We analyzed the correlation between the features and experimentally measured LMPs of AFA alloys with two different methods: conventional Pearson's correlation coefficient (PCC) and advanced maximal information coefficient (MIC) [23]. While the Pearson correlation analysis only seeks for linear-relationships between input features and target properties, MIC can identify non-linear relationships of high-dimensional large dataset [23]. Pearson analysis can have negative values when the correlation is inverse. Hence, the squared values of Pearson correlation analysis (PCC$^2$) have been used to compare the scores obtained from the MIC analysis for the comparison. The significance of correlation analysis is two-fold: (1) select features to be included in the training of machine learning models, and (2) generate alloy hypotheses to reveal underlying mechanisms and to guide developing future strategies for improved alloy design.

Selecting features for the training of machine learning models is a multi-variant regression/optimization problem, and thus it is critical to have a numerical/statistical basis. Both Pearson and MIC approaches can provide rankings based on quantitative analysis to determine a threshold score for feature selection. The other and more critical aspect of correlation analysis is the validation/generation of materials hypotheses [24]. If highly-ranked features are already known and generally accepted, then correlation analysis can confirm the already established mechanisms. On the other hand, if overlooked/hidden features are highly ranked over conventionally anticipated features, then these hidden correlations can potentially enable generation of new materials hypotheses. It should be emphasized here that correlation analysis approaches do not heuristically suggest new features that are anticipated to be better than existing ones, but only rank the provided feature set. Hence, it is critically important to incorporate input from materials domain experts to compile an initial advanced feature list and then iteratively improve it based on the correlation analysis results. Herein, we have started with the initial alloy thermodynamic features that are closely related to microstructure of AFA alloys, but the approach shown here can be expanded to accommodate feature group from other alloy physics, such as diffusion kinetics and microstructure evolution.

2.4. Training of machine learning models

We have used five different representative machine learning models in the present study: random forest (RF), linear regression (LR), nearest neighbor (NN), kernel ridge (KR), Bayesian ridge (BR) as implemented in the Python-based open source data analytics toolkit, *scikit-learn* [25]. The individual models are explained in detail elsewhere [1]. The present work intends to demonstrate a workflow that integrates highly relevant synthetic/scientific data with rigorously consistent, high-quality experimental data within the context of data analytics. Hence, we report the performance of different machine learning models as a function of the number of features used in training instead of reporting optimum number of features for a specific machine learning model. We have used default values for hyperparameters of each machine learning model and the *k*-fold cross validation scheme (*k*=5). We trained each machine learning model at the given number of features 10 times to get an averaged accuracy, represented as Pearson correlation coefficient (PCC).

## 3. Results and Discussion

### 3.1. Correlation analysis

#### 3.1.1. Simple/superficial features: compositions only

We first analyzed the correlation between simple/superficial features of AFA alloys, i.e., bulk alloy analyzed elemental compositions (at%), creep stress and LMPs with Pearson and MIC methods. In addition, we included an engineered feature, the at% elemental ratio between compositions of Nb and C (denoted as NB_C). This engineered Nb_C feature was selected because the longest creep-rupture lives of a select subset of AFA alloys at 750°C 100 and 170 MPa testing was observed in alloys with near stoichiometric amounts of Nb and C additions (0.6~1.0 and 0.1 wt%, respectively) [19]. The rankings and scores of features from Pearson and MIC analyses are summarized in Figure 3 and Table 3.

[Figure 3 about here]

[Table 3 about here]

As shown in Figure 3 and Table 3, both Pearson and MIC analyses respectively identify stress and T2 as the strongest feature that affects LMP of AFA alloys in the given alloy compositions and vice versa for the second strongest. This finding is promising from a data analytics viewpoint: both stress and T2 are obviously key factors in determining creep rupture life, and the correlation analysis correctly identified this. Further, in the case of the Pearson analysis, stress was shown to be a strong negative factor in LMP. As such, there is some reassurance that other descriptors with high ranking correlations may provide reliable insights into alloy-LMP relationships.

Interestingly, the five highly ranked features from Pearson and MIC analyses are identical (stress, T2, dT, xMN, and xCU), but there is a discrepancy between the two methods thereafter. For example, MIC ranked xNB as 6[th] (17[th] in Pearson), but Pearson ranked xNI as 6[th] (16[th] in MIC). It should be noted here that the purpose of comparing two different correlation analysis methods is not to identify a superior approach over the other. Correlation analysis is a topic of its own where its objective is to study the statistical relationship strength between two continuous variables, in this case, a superficial feature and LMP. There are a number of correlation analysis techniques such as rank correlation coefficient [26], distance correlation coefficient [27], randomized dependence coefficient [28], polychoric correlation coefficient [29], etc. Correlation analysis techniques can be considered as a category of feature selection approaches, which aim to select a subset of relevant/useful features for use in machine learning models. There are feature selection approaches such as mutual information [30], relief algorithms [31,32], and stepwise regression [33] that share the similar goal of our approach in this section. Each technique captures different statistical relationships.

Rather, we wish to demonstrate that the results of different correlation analyses, PCC and MIC, can be further analyzed to inspire domain experts for generating alloy hypotheses. Along this line, both elemental compositions of Nb and Ni ranked 6[th] in Pearson and MIC, respectively, can be associated with experimentally suggested strengthening mechanisms of AFA alloys [19,20]: Nb favors the formation of NbC and $Fe_2Nb$, and Ni stabilizes austenite Fe matrix, B2-NiAl, and $L1_2$-$Ni_3Al$, respectively. However, individual elemental compositions can only indirectly explain the underlying strengthening mechanisms of AFA alloys and does not provide a basis for a more fundamental insight into why they impact LMP. As such, they are of limited utility, and extensions to prediction of new alloy compositions as well without the feature(s) can be used to generate hypotheses. The engineered Nb_C feature was not highly correlated, likely in part because this feature peaks for LMP benefit near a stoichiometric Nb:C ratio related to NbC strengthening precipitates, and then declines. Therefore, it is critical to incorporate advanced scientific features that can more directly be associated with target behavior and properties. In this case, calculated phase stabilities of key precipitates, represented as volume/phase fractions and individual phase elemental compositions/activities at given temperatures, for example, may be more closely correlated with alloy creep properties than simple bulk alloy compositions.

### 3.1.2. Advanced synthetic/scientific features

We next analyzed the relationship between advanced scientific features populated from equilibrium thermodynamic calculations of AFA alloys and LMP. We have considered these calculated features in addition to simple/superficial features discussed in the previous section. The total number of features, both simple and scientific, is 466. The correlation analysis results of the top 30 features from Pearson and MIC are summarized in Figure 4 and Table 4. The interpretation of correlation analyses results can be highly sensitive to the threshold ranking. We chose to closely investigate top 30 features. The complete list of rankings for all features from both Pearson and MIC analyses are provided in the supplementary information.

[Figure 4 about here]

[Table 4 about here]

It was encouraging that both analysis methods again recognized stress and T2 as the top-ranking features among considered ones. Beyond two top-ranked stress and T2, Pearson and MIC ranked a somewhat similar group of features in the top 30. Overall, MIC assigns high rankings to NbC, NiAl, FCC, M2B phases-related features, while Pearson highly ranks FCC, NbC, sigma, and M2B phase-related features. Both analyses also gave higher scores to calculated features computed at the creep testing temperature (T2) rather than the solutionizing heat treatment temperature (T1). Intriguingly, although the original AFA alloy design strategy was based on maximizing phase fractions and degree of supersaturation from T1 to T2 of key strengthening phases, NbC, $M_{23}C_6$, and in selected alloys $L1_2$, the highest-ranking features in both Pearson and MIC were dominated by individual phase elemental composition and activity features. This finding suggests more consideration of strengthening phase composition/activity design aspects in future AFA alloy development efforts, as well as revisiting select AFA alloys to focus more experimental characterization effort on measurement of phase compositions per highly ranked Pearson and MIC features. It is speculated that the high rankings of these type of features may be secondary measures of kinetic factors related to coarsening behavior, suggesting future data analytics in this dataset be populated with as many diffusion kinetics and microstructure morphological features as feasible.

## 3.2. Training of machine learning models

Based on the rankings of features obtained from MIC and Pearson correlation analyses, we performed two rounds of machine learning trainings: with and without calculated scientific alloy features populated from high-throughput computational thermodynamics. Instead of optimizing the number of features for five respective machine learning models, we wish to investigate the performance of each model by varying the number of features used within. The training results of five machine learning models with and without advanced scientific features with different number of features based on the respective rankings from MIC and Pearson analyses are presented in Figures 5 and 6, respectively.

[Figure 5 about here]

Figure 5 shows that the overall accuracy of all the machine learning models, except kernel ridge, is above 90%. In contrast to anticipated outcomes, this was true even with simple/superficial features. The highest accuracy was obtained with random forest using the top 10 features from MIC. An increase and decrease of machine learning accuracy can be interpreted with respect to noise and key features. For example, excluding physically meaningful key features should yield a decrease in the accuracy of machine learning model. Similarly, rejecting noise (i.e., physically non-relevant features) is anticipated to increase in the accuracy of surrogate models.

It is remarkable that relatively high accuracy (above 90%) was demonstrated in the training of machine learning models with only simple/superficial features, i.e., bulk alloy elemental compositions, and stress (see Table 3 for individual features and their ranking from Pearson and MIC). We believe that consistency in the experimental AFA data, i.e., all AFA alloys manufactured and hot/cold worked in a similar manner and tested as-solutionized, contributed to the high accuracy of machine learning models with small error bars. However, even with such high accuracy in predicting LMPs only with bulk alloy chemistry and stress as an input, this approach incorporating only simple/superficial features in training machine learning models is equivalent to developing a 'black box', with very limited insight for understanding the underlying creep behavior and generating new alloy design approaches to improve the alloys. Hence, although this approach can be used to predict creep properties of AFA alloys within the composition range of individual bulk alloying elements, interrogation of developed surrogate machine learning models will rely on exhaustive numerical enumeration of hypothetical AFA alloys.

Next, we trained machine learning models with all the calculated scientific features populated from the high-throughput computational thermodynamics calculations. Similarly, we varied the number of features based on the MIC and Pearson analyses for training of each machine learning model. Accuracy (represented as PCC) of five machine learning models with respect to the number of top-ranking features are presented in Figure 6 Tranining results of machine learning models with synthetic/scientific features populated from high-throughput computational thermodynamic approach: (a) accuracy, represented as PCC of five different machine learning models (RF: random forest, LR: linear regression, NN: nearest neighbor, KR: kernel ridge regression, BR: Bayesian ridge regression) as a function of number of top ranking features, (b) actual vs predicted LMP with random forest with top 50 features, (c) actual vs predicted LMP with linear regression with top 300 features.

[Figure 6 about here]

Overall, random forest, neural network, and Bayesian ridge models exhibit high accuracy (~90%) regardless of the number of features used within the training. These three models achieved similar accuracy at any given number of features from MIC and Pearson analyses. On the contrary, linear regression is highly sensitive to the number of features used in the training and it exhibits exactly the opposite trend: the accuracy of training gets better as the number of feature decreases.

The accuracy of trained machine learning models with calculated scientific features are in general similar to those trained only with simple/superficial features. Further improvement can be made if hyperparameters within individual machine learning models are further optimized. However, such efforts may increase the accuracy of the prediction, without increasing scientific knowledge to better understand underlying mechanisms and providing insights for new AFA alloy design. A more meaningful approach to improve the accuracy will be incorporating features from factors beyond calculated thermodynamic phase equilibria that play an important role affect creep, e.g., diffusion kinetics (atomic mobility of individual elements) and morphological microstructure information (such as size, shape and distribution of precipitates, etc). Such an attempt is anticipated to bring opportunities to consider features at different length scale concurrently, and modern data analytics will serve as an excellent wrapping planform to realize high-level knowledge integration. Input from domain experts again will play a critical to efficiently create a feature strategy from different aspect of creep of high-temperature alloys.

### 3.3. Prediction from trained machine learning models

As a final step, we wish to predict LMPs of hypothetical AFA alloys, of which compositions have not been used in the training of machine learning model. We arbitrarily generated compositions of AFA alloys within the ranges of individual alloying elements and their predicted LMPs are selectively plotted along with those of used in the training of machine learning models in Figure 7. We have only used the machine learning models trained with simple/superficial composition for a demonstration purpose. The ultimate goal of exploiting trained machine learning models would be identifying alloy compositions and phases/microstructures chemistry, of which predicted LMP is larger than the experimentally measured ones at a given stress.

[Figure 7 about here]

However, as the number of elements in AFA alloys considered in the present work is 16, considering $n$ number of variations in each element will require $n^{16}$ possible combination. It is possible to apply certain level of constraints to reduce the composition search space, or one can use advanced global minimum search algorithms to more efficiently exploit the developed machine learning models. However, solely relying on machine learning models, even the ones trained with high-quality experimental data augmented with scientific features, to optimize alloy chemistry to design advanced high-temperature alloys at present may be premature. Instead, it would be more desirable for physical metallurgists/alloy designers to exercise this newly developed workflow to reduce the number of prototype alloys that they fabricate, test and characterize by proving them with highly ranked scientific features selected from a large set of possibilities, beyond the capability of human analysis alone (e.g., 466 phase equilibria related features in the present example). The modern data analytics approach that combines correlation analysis and machine learning should be regarded as additional tools that facilitate and accelerate understanding of underlying mechanism to guide alloy design.

## 4. Conclusion

We introduce a modern data analytics workflow that has potential to reduce the number of prototype samples for experimental validation by leveraging consistently measured high-quality data and relevant large synthetic datasets of selected scientific features. We use an example of the creep resistance represented as LMP of developmental AFA alloys, of which data have been collected over a decade and detailed metadata (e.g., analyzed alloy composition and processing history) is available. We go beyond legacy data analytics approaches that only correlate simple/superficial features (e.g., bulk alloy elemental compositions and processing temperature) with alloy properties by augmenting raw data with advanced scientific features populated from high-fidelity materials models. We have interrogated state-of-the-art computational thermodynamic databases to compute microstructure-related alloy features, such as phase/volume fraction, chemistry, and degree of supersaturation, in a high-throughput manner. We then performed extensive correlation analyses to identify key features that affect LMP of AFA alloys and confirmed that highly ranked features concur with generally accepted strengthening mechanisms. Some additional insights regarding the potential importance of individual phase chemistries not considered in the original AFA alloy design effort have been also obtained.

Intriguingly, incorporating scientific features did not greatly improve the accuracy in the training of machine learning models to predict LMP as it was already above 90% with only simple/superficial features. We believe this outcome can be attributed to the consistently measured high-quality experimental data with known metadata. Although the demonstrated data analytics approach that combines correlation analysis and machine learning may offer significant opportunities to improve high-temperature alloy design in the near future, it should be regarded as an intermediate tool to facilitate new alloy design insights and approaches to guide prototype alloy selection and experimental validation. Finally, we believe the current approach should be expanded to accommodate additional important scientific features other than thermodynamics-based ones, such as those related to diffusion kinetics, coarsening, and microstructure morphology (e.g., precipitate size, shape, and distribution). Perhaps it will be possible to include synthetic microstructures generated from phase-field simulations in the foreseeable future. The role of domain experts again will be critical to create a feature strategy that concurrently considers the most relevant feature set at different alloy physics and length-scales.

**Acknowledgement**

This research was sponsored by the Laboratory Directed Research and Development Program of ORNL, managed by UT-Battelle, LLC, for the US Department of Energy. DS would like to thank Turab Lookman and John Vitek for their fruitful discussion.

**References**

[1] D. Shin, S. Lee, A. Shyam, J.A. Haynes, Petascale supercomputing to accelerate the design of high-temperature alloys, Sci. Technol. Adv. Mater. 18 (2017) 828–838. doi:10.1080/14686996.2017.1371559.

[2] Y.K. Kim, D. Kim, H.K. Kim, C.S. Oh, B.J. Lee, An intermediate temperature creep model for Ni-based superalloys, Int. J. Plast. 79 (2016) 153–175. doi:10.1016/j.ijplas.2015.12.008.

[3] F. Tancret, H.K.D.H. Bhadeshia, D.J.C. MacKay, Design of a creep resistant nickel base superalloy for power plant applications: Part 1 - Mechanical properties modelling, Mater. Sci. Technol. 19 (2003) 283–290. doi:10.1179/026708303225009788.

[4] F. Brun, T. Yoshida, J.D. Robson, V. Narayan, H.K.D.H. Bhadeshia, D.J.C. MacKay, Theoretical design of ferritic creep resistant steels using neural network, kinetic, and thermodynamic models, Mater. Sci. Technol. 15 (1999) 547–554. doi:10.1179/026708399101506085.

[5] H.K.D.H. Bhadeshia, T. Sourmail, Design of Creep – Resistant Steels : Success & Failure of Models, Japan Soc. Promot. Sci. Comm. Heat–Resisting Mater. Alloy. 44 (2003) 299–314.

[6] R.C. Dimitriu, H.K.D.H. Bhadeshia, Hot strength of creep resistant ferritic steels and relationship to creep rupture data, Mater. Sci. Technol. 23 (2007) 1127–1131. doi:10.1109/DSN.2017.64.

[7] H.K.D.H. Bhadeshia, Neural networks in materials science, ISIJ Int. 39 (1999) 966–979. doi:10.2355/isijinternational.39.966.

[8] L. Gavard, H.K.D.H. Bhadeshia, D.J.C. MacKay, S. Suzuki, Bayesian neural network model for austenite formation in steels, Mater. Sci. Technol. 12 (1996) 453–463. doi:10.1179/mst.1996.12.6.453.

[9] T. Sourmail, H.K.D.H. Bhadeshia, D.J.C. MacKay, Neural network model of creep strength of austenitic stainless steels, Mater. Sci. Technol. 18 (2002) 655–663. doi:10.1179/026708302225002065.

[10] A. Agrawal, P.D. Deshpande, A. Cecen, G.P. Basavarsu, A.N. Choudhary, S.R. Kalidindi, Exploration of data science techniques to predict fatigue strength of steel from composition and processing parameters, Integr. Mater. Manuf. Innov. 3 (2014) 8. doi:10.1186/2193-9772-3-8.

[11] L. Ward, A. Agrawal, A. Choudhary, C. Wolverton, A General-Purpose Machine Learning Framework for Predicting Properties of Inorganic Materials, Nat. Commun. (2015) 1–7. doi:10.1038/npjcompumats.2016.28.

[12] L. Ward, S. O'Keeffe, J. Stevik, G.R. Jelbert, M. Aykol, C. Wolverton, A Machine Learning Approach for Engineering Bulk Metallic Glass Alloys, Mater. Data Facil. 159 (2018) 102–111. doi:10.18126/M2662X.

[13] L. Ward, S.C. O'Keeffe, J. Stevick, G.R. Jelbert, M. Aykol, C. Wolverton, A machine learning approach for engineering bulk metallic glass alloys, Acta Mater. 159 (2018) 102–111. doi:10.1016/j.actamat.2018.08.002.

[14] Y. Yamamoto, M.P. Brady, Z.P. Lu, P.J. Maziasz, C.T. Liu, B.A. Pint, K.L. More, H.M. Meyer, E.A. Payzant, Creep-Resistant, Al2O3-Forming Austenitic Stainless Steels, Science

(80-. ). 316 (2007) 433–436. doi:10.1126/science.1137711.

[15]  M.P. Brady, Y. Yamamoto, M.L. Santella, P.J. Maziasz, B.A. Pint, C.T. Liu, Z.P. Lu, H. Bei, The development of alumina-forming austenitic stainless steels for high-temperature structural use, JOM. 60 (2008) 12–18. doi:10.1007/s11837-008-0083-2.

[16]  M.P. Brady, K.A. Unocic, M.J. Lance, M.L. Santella, Y. Yamamoto, L.R. Walker, Increasing the Upper Temperature Oxidation Limit of Alumina Forming Austenitic Stainless Steels in Air with Water Vapor, Oxid. Met. 75 (2011) 337–357. doi:10.1007/s11085-011-9237-7.

[17]  M.P. Brady, J. Magee, Y. Yamamoto, D. Helmick, L. Wang, Co-optimization of wrought alumina-forming austenitic stainless steel composition ranges for high-temperature creep and oxidation/corrosion resistance, Mater. Sci. Eng. A. 590 (2014) 101–115. doi:10.1016/j.msea.2013.10.014.

[18]  B.A. Pint, S. Dryepondt, M.P. Brady, Y. Yamamoto, B. Ruan, R.D. McKeirnan, Field and Laboratory Evaluations of Commercial and Next-Generation Alumina-Forming Austenitic Foil for Advanced Recuperators, J. Eng. Gas Turbines Power. 138 (2016) 1–5. doi:10.1115/1.4033746.

[19]  Y. Yamamoto, M.P. Brady, M.L. Santella, H. Bei, P.J. Maziasz, B.A. Pint, Overview of strategies for high-temperature creep and oxidation resistance of alumina-forming austenitic stainless steels, Metall. Mater. Trans. A Phys. Metall. Mater. Sci. 42 (2011) 922–931. doi:10.1007/s11661-010-0295-2.

[20]  Y. Yamamoto, M.L. Santella, M.P. Brady, H. Bei, P.J. Maziasz, Effect of alloying additions on phase equilibria and creep resistance of alumina-forming austenitic stainless steels, Metall. Mater. Trans. A Phys. Metall. Mater. Sci. 40 (2009) 1868–1880. doi:10.1007/s11661-009-9886-1.

[21]  Y. Yamamoto, M.P. Brady, Z.P. Lu, P.J. Maziasz, C.T. Liu, B.A. Pint, K.L. More, H.M. Meyer, E.A. Payzant, Creep-resistant, Al2O3-forming austenitic stainless steels, Science (80-. ). 316 (2007) 433–436.

[22]  N. Saunders, A.P. Miodownik, CALPHAD (Calculation of Phase Diagrams): A Comprehensive Guide, Pergamon, Oxford, New York, 1998.

[23]  D.N. Reshef, Y.A. Reshef, H.K. Finucane, S.R. Grossman, G. McVean, P.J. Turnbaugh, E.S. Lander, M. Mitzenmacher, P.C. Sabeti, Detecting Novel Associations in Large Data Sets, Science (80-. ). 334 (2011) 1518–1524. doi:10.1126/science.1205438.

[24]  S. Ramakrishna, T.-Y. Zhang, W.-C. Lu, Q. Qian, J.S.C. Low, J.H.R. Yune, D.Z.L. Tan, S. Bressan, S. Sanvito, S.R. Kalidindi, Materials informatics, J. Intell. Manuf. (2018). doi:10.1007/s10845-018-1392-0.

[25]  F. Pedregosa, G. Varoquaux, A. Gramfort, V. Michel, B. Thirion, O. Grisel, M. Blondel, G. Louppe, P. Prettenhofer, R. Weiss, V. Dubourg, J. Vanderplas, A. Passos, D. Cournapeau, M. Brucher, M. Perrot, É. Duchesnay, Scikit-learn: Machine Learning in Python, 12 (2012) 2825–2830. doi:10.1007/s13398-014-0173-7.2.

[26]  M.G. Kendall, A New Measure of Rank Correlation, Biometrika. 30 (1938) 81. doi:10.2307/2332226.

[27]  G.J. Székely, M.L. Rizzo, N.K. Bakirov, Measuring and testing dependence by correlation of distances, Ann. Stat. 35 (2007) 2769–2794. doi:10.1214/009053607000000505.

[28] D. Lopez-Paz, P. Hennig, B. Schölkopf, The Randomized Dependence Coefficient, in: C.J.C. Burges, L. Bottou, M. Welling, Z. Ghahramani, K. Weinberger, Q. (Eds.), Adv. Neural Inf. Process. Syst. 26, Curran Associates, Inc., 2013: pp. 1–9. doi:10.1214/aos/1176345528.

[29] U. Olsson, Maximum Likelihood Estimation of the Polychoric, 4 (1979) 443–460.

[30] I. Guyon, A. Elisseeff, An Introduction to Variable and Feature Selection, J. Mach. Learn. Res. 3 (2003) 1157–1182. doi:10.1016/j.aca.2011.07.027.

[31] K. Kira, L. Rendell, The feature selection problem: Traditional methods and a new algorithm, Aaai. (1992) 129–134. doi:10.1016/S0031-3203(01)00046-2.

[32] I. Kononenko, E. Simec, M.R.- Sikonja, Overcoming the Myopia of Inductive Learning Algorithms with RELIEFF, Appl. Intell. 7 (1997) 39–55. doi:10.1023/A:1008280620621.

[33] B.B. Robert, A.A. Afifi, Comparison of Stopping Rules in Forward "Stepwise" Regression, J. Am. Stat. Assoc. 72 (1977) 46–53.

**List of Tables**

Table 1 List of considered elements with composition range in wt%, temperature, stress and phases of the AFA experimental creep dataset and computational thermodynamic calculations.

Table 2 List of considered alloy features (i.e., simple/superficial and computed scientific) used in the present work.

Table 3 Correlation analysis results between simple/superficial features (elemental compositions, stress, temperature difference between heat treatment and operation, and the ratio between Nb and C) and LMP of AFA alloys using MIC and Pearson methods.

Table 4 Correlation analysis results between synthetic/scientific features populated from high-throughput computational thermodynamic approach and LMP of AFA alloys using MIC and Pearson methods. Only top 30 features are shown. Pearson feature ranking with a negative impact on LMP is presented in parenthesis.

Table 1 List of considered elements with composition range in wt%, temperature, stress and phases of the AFA experimental creep dataset and computational thermodynamic calculations.

| | | |
|---|---|---|
| Elements | | Fe (Bal.), Ni (12-32), Cr (12-20), Al (0-5), Nb (0-3.3), Ti (0-1), V (0-1), Mo (0-2), W (0-2), Si (0-1), Mn (0-12), Cu (0-3), Zr (0-0.5), Y(0-0.2), C (0-0.2), B (0-0.1) |
| T1 (solutionizing, °C) | | 1100, 1150, 1200, 1250 |
| T2 (creep test, °C) | | 650, 700, 750, 800 |
| Stress (MPa) | | 70, 100, 130, 170, 200, 250, 300 |
| Phases | FCC | Austenite (FCC matrix), NbC, L1$_2$ |
| | BCC (B2) | (Fe,Ni)Al |
| | C14 Laves | Fe$_2$(Mo,Nb) |
| | others | M$_{23}$C$_6$, M$_2$B, M$_3$B$_2$, MB$_2$, Nb$_3$B$_2$, NbNi$_3$, Sigma |

Table 2 List of considered alloy features (i.e., simple/superficial and computed scientific) used in the present work.

| Features | | Descriptions |
|---|---|---|
| Simple/ superficial | Stress | Creep stress |
| | | Solutionizing, creep test temperatures, temperature difference between T1 and T2 |
| | NB_C[‡] | Ratio between Nb and C |
| | x{Elements} | Elemental composition (at %) |
| Synthetic/ scientific | {Temperature}_NPV_{Phases} | Phase fraction at T1 and T2 |
| | {Temperature}_VPV_{Phases} | Volume fraction at T1 and T2 |
| | {Temperature}_ACR_{Phases}_{Elements} | Activity of elements in phases at T1 and T2 |
| | {Temperature}_{Phases}_X_{Elements} | Composition of elements in phases at T1 and T2 |
| | {Temperature}_{Phases}_CX_{Elements}[‡] | {Phase fraction} × {elemental composition} |
| | d{Phases}[‡] | Degree of supersaturation (difference in volume fraction of each phase between T1 and T2) |

[†]Not included in the analysis as it is embedded in LMP

[‡]Engineered features

Table 3 Correlation analysis results between simple/superficial features (elemental compositions, stress, temperature difference between heat treatment and operation, and the ratio between Nb and C) and LMP of AFA alloys using MIC and Pearson methods.

| Features | MIC strength | PCC$^2$ | PCC | MIC Ranking | Pearson Ranking |
|---|---|---|---|---|---|
| T2 | 0.775 | 0.871 | 0.871 | 1 | 2 |
| Stress | 0.735 | 0.898 | -0.898 | 2 | 1 |
| dT | 0.570 | 0.682 | -0.682 | 3 | 3 |
| xMN | 0.333 | 0.256 | -0.256 | 4 | 4 |
| xCU | 0.320 | 0.252 | -0.252 | 5 | 5 |
| xNB | 0.314 | 0.051 | 0.051 | 6 | 17 |
| xY | 0.296 | 0.056 | 0.056 | 7 | 16 |
| xMO | 0.293 | 0.080 | 0.080 | 8 | 13 |
| Nb_C | 0.286 | 0.011 | 0.011 | 9 | 20 |
| xZR | 0.283 | 0.194 | 0.194 | 10 | 7 |
| xC | 0.275 | 0.069 | -0.069 | 11 | 15 |
| xB | 0.253 | 0.008 | 0.008 | 12 | 21 |
| xSI | 0.246 | 0.081 | 0.081 | 13 | 12 |
| xW | 0.241 | 0.076 | 0.076 | 14 | 14 |
| xAL | 0.233 | 0.090 | 0.090 | 15 | 11 |
| xNI | 0.229 | 0.219 | 0.219 | 16 | 6 |
| xCR | 0.216 | 0.019 | -0.019 | 17 | 19 |
| T1 | 0.209 | 0.023 | -0.023 | 18 | 18 |
| xV | 0.204 | 0.160 | -0.160 | 19 | 10 |
| xFE | 0.183 | 0.162 | -0.162 | 20 | 9 |
| xTI | 0.182 | 0.185 | 0.185 | 21 | 8 |

Table 4 Correlation analysis results between synthetic/scientific features populated from high-throughput computational thermodynamic approach and LMP of AFA alloys using MIC and Pearson methods. Only top 30 features are shown. Pearson feature ranking with a negative impact on LMP is presented in parenthesis.

| Features | MIC Ranking | Pearson Ranking | Features | Pearson Ranking | MIC Ranking |
|---|---|---|---|---|---|
| T2 | 1 | 2 | Stress | (1) | 2 |
| Stress | 2 | 1 | T2 | 2 | 1 |
| T2_NbC_ACR_AL | 3 | 5 | T2_FCC_ACR_AL | 3 | 12 |
| T2_NIAL_ACR_B2_AL | 4 | 86 | dT | (4) | 10 |
| T2_NbC_ACR_SI | 5 | 124 | T2_NbC_ACR_AL | 5 | 3 |
| T2_FCC_ACR_SI | 6 | 104 | T2_SIGMA_X_V | (6) | 121 |
| T2_NIAL_B2_X_FE | 7 | 157 | T2_SIGMA_X_W | (7) | 122 |
| T2_NIAL_B2_X_NI | 8 | 76 | T2_SIGMA_X_FE | (8) | 116 |
| T2_NbC_X_AL | 9 | 9 | T2_NbC_X_AL | 9 | 9 |
| dT | 10 | 4 | T2_M2B_CB_X_CR | (10) | 32 |
| T2_FCC_X_B | 11 | 94 | T2_M2B_CB_CX_MO | (11) | 29 |
| T2_FCC_ACR_AL | 12 | 3 | T2_M2B_CB_X_B | (12) | 31 |
| T2_FCC_CX_B | 13 | 92 | T2_NbC_CX_AL | 13 | 14 |
| T2_NbC_CR_AL | 14 | 13 | T2_SIGMA_X_CR | (14) | 115 |
| T2_NbC_X_NI | 15 | 30 | T2_M2B_CB_X_MO | (15) | 35 |
| T2_NbC_CX_B | 16 | 120 | T2_NbC_X_MO | 16 | 50 |
| T2_NbC_X_B | 17 | 128 | T2_M2B_CB_X_FE | (17) | 33 |
| T2_NIAL_ACR_B2_FE | 18 | 47 | T2_NbC_CX_MO | 18 | 39 |
| T2_NIAL_B2_X_CR | 19 | 358 | T2_M3B2_D5A_X_NI | 19 | 72 |
| T2_M2B_CB_X_NI | 20 | 208 | T2_SIGMA_X_AL | (20) | 114 |
| T2_FCC_ACR_MO | 21 | 61 | T2_SIGMA_X_MO | (21) | 117 |
| T2_NbC_ACR_MO | 22 | 62 | T2_LAVES_C14_X_AL | 22 | 23 |
| T2_LAVES_C14_X_AL | 23 | 22 | T2_VPV_M2B_CB | (23) | 37 |
| dM2B_CB | 24 | 32 | T2_M2B_CB_CX_CR | (24) | 26 |
| T2_M2B_CB_CX_B | 25 | 25 | T2_M2B_CB_CX_B | (25) | 25 |
| T2_M2B_CB_CX_CR | 26 | 24 | T2_NP_M2B_CB | (26) | 36 |
| T2_M2B_CB_CX_FE | 27 | 31 | T2_LAVES_C14_ACR_AL | 27 | 51 |
| T2_M2B_CB_CX_MN | 28 | 79 | T2_NbC_CX_NI | 28 | 44 |
| T2_M2B_CB_CX_MO | 29 | 11 | T2_SIGMA_X_MN | (29) | 99 |
| T2_M2B_CB_CX_NI | 30 | 213 | T2_NbC_X_NI | 30 | 15 |

**List of Figures**

Figure 1 Typical microstructure in AFA alloys; (a) SEM backscattered electron images of Fe-12Cr-4Al-20Ni base AFA alloy after aging at 750°C for 2,000h, and (b) TEM bright field image of Fe-14Cr-2.5Al-20Ni base AFA alloy after creep testing of at 750°C and 100 MPa for 2,000 h [19]. Other AFA alloy variations use M23C6 or γ'-Ni3Al strengthening precipitates.

Figure 2 Experimentally measured Larson-Miller Parameter (LMP) of AFA alloys plotted with respect to creep stress (open circles) and the ones that have been used in the present study (closed circles). Data from the cold-worked samples have been excluded as they have different strengthening phase precipitation behaviors. Rather, only data involving as-solutionized microstructures were used for data analytics.

Figure 3 Correlation analysis between superficial features and LMP of AFA alloys from (a) Pearson correlation represented as PCC, and (b) from MIC. The MIC result has been compared with PCC2.

Figure 4 Correlation analysis between all the alloy features (simple/superficial and microstructure-related synthetic/scientific features generated from high-throughput Thermo-Calc computation) and LMP of AFA alloys from (a) Pearson correlation represented as PCC, and (b) from MIC. The MIC result has been compared with PCC2. Top 15 and bottom 15 ranking features out of 466 features are shown in (a), and only top 30 ranking features from MIC analysis and correpsonding features from Pearson correlation analysis are shown in (b), respectively.

Figure 5 Training results of machine learning models with simple/superficial features, i.e., elemental compositions, stress, solutionizing temperature, and the ratio between Nb and C: (a) accuracy, represented as PCC of five different machine learning models (RF: random forest, LR: linear regression, NN: nearest neighbor, KR: kernel ridge regression, BR: Bayesian ridge regression) as a function of number of top ranking features, (b) actual vs predicted LMP with random forest with top 5 features, (c) actual vs predicted LMP with kernel ridge regression with top 15 features.

Figure 6 Tranining results of machine learning models with synthetic/scientific features populated from high-throughput computational thermodynamic approach: (a) accuracy, represented as PCC of five different machine learning models (RF: random forest, LR: linear regression, NN: nearest neighbor, KR: kernel ridge regression, BR: Bayesian ridge regression) as a function of number of top ranking features, (b) actual vs predicted LMP with random forest with top 50 features, (c) actual vs predicted LMP with linear regression with top 300 features.

Figure 7 Experimentally measured LMPs of AFA alloys used for the training of machine learning models and predicted LMPs of hypothetical AFA alloys from trained models.

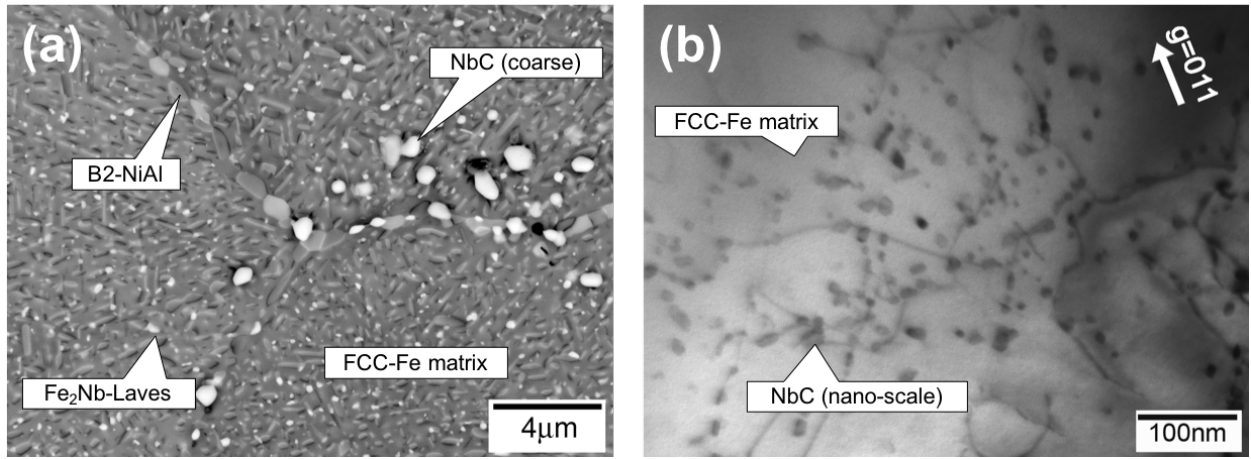

Figure 1 Typical microstructure in AFA alloys; (a) SEM backscattered electron images of Fe-12Cr-4Al-20Ni base AFA alloy after aging at 750°C for 2,000h, and (b) TEM bright field image of Fe-14Cr-2.5Al-20Ni base AFA alloy after creep testing of at 750°C and 100 MPa for 2,000 h [19]. Other AFA alloy variations use $M_{23}C_6$ or $\gamma'$-$Ni_3Al$ strengthening precipitates.

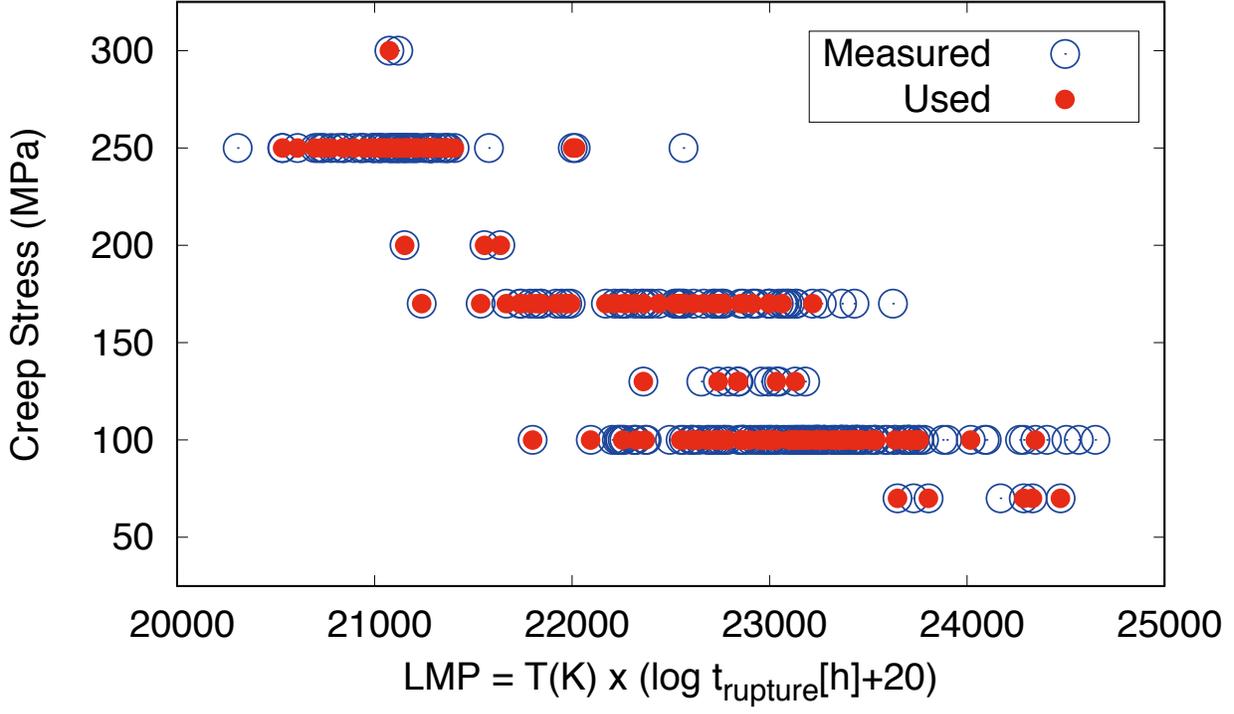

Figure 2 Experimentally measured Larson-Miller Parameter (LMP) of AFA alloys plotted with respect to creep stress (open circles) and the ones that have been used in the present study (closed circles). Data from the cold-worked samples have been excluded as they have different strengthening phase precipitation behaviors. Rather, only data involving as-solutionized microstructures were used for data analytics.

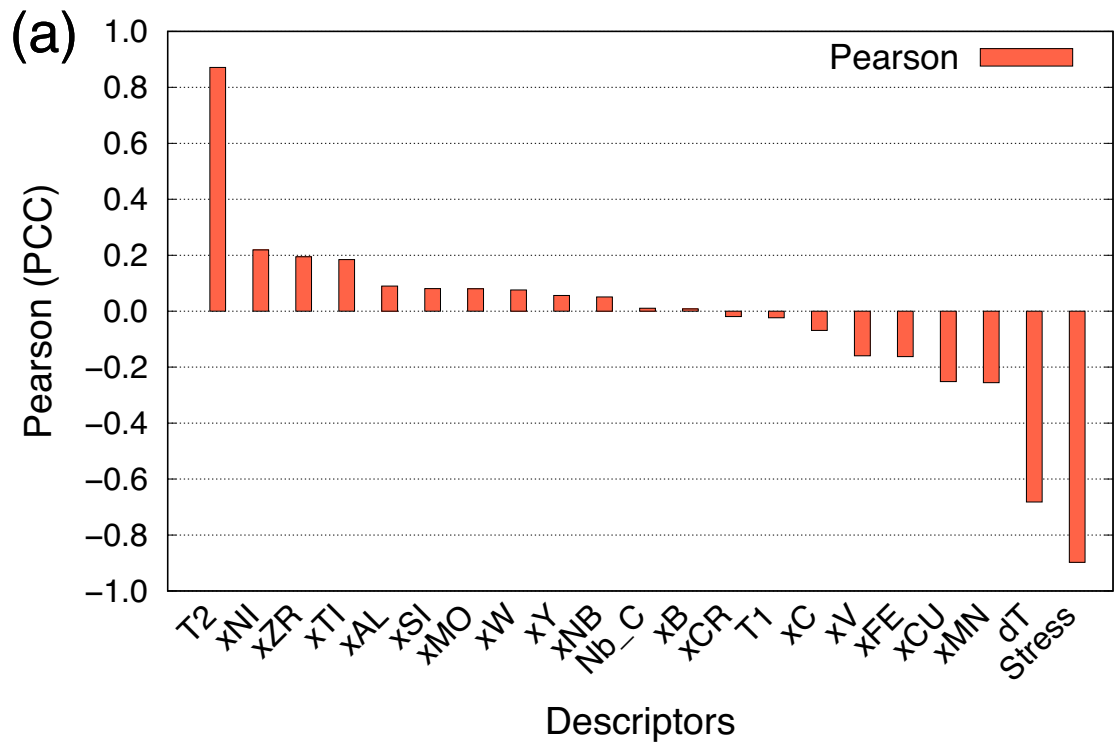

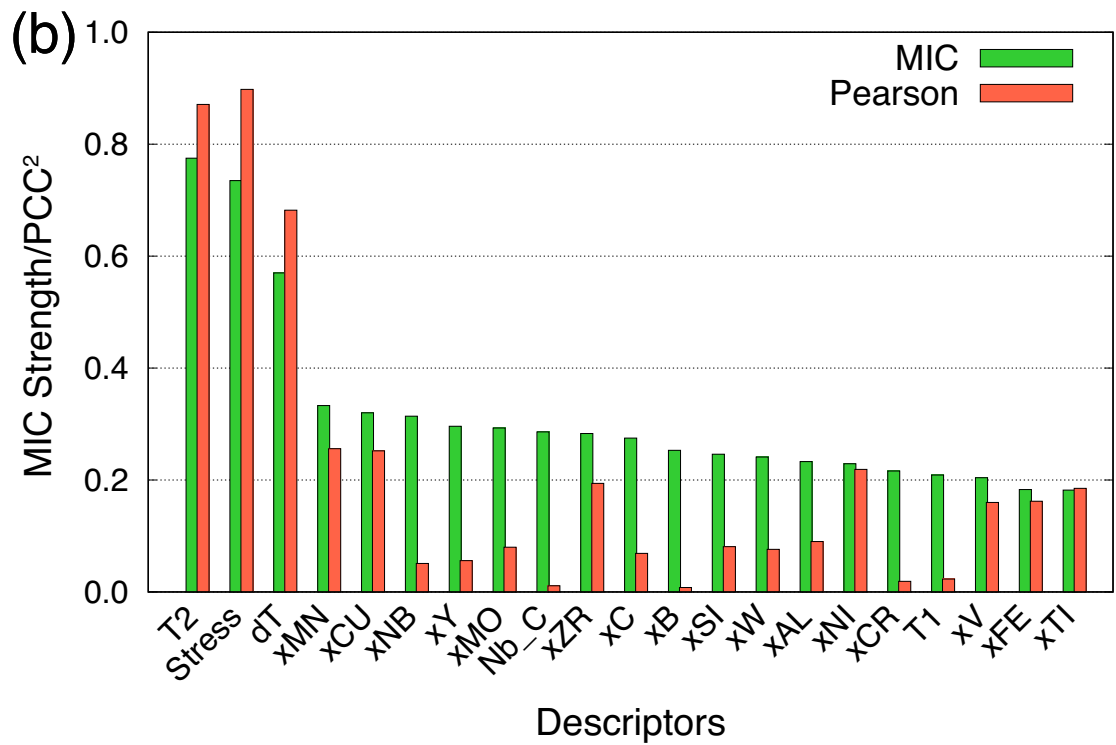

Figure 3 Correlation analysis between superficial features and LMP of AFA alloys from (a) Pearson correlation represented as PCC, and (b) from MIC. The MIC result has been compared with $PCC^2$.

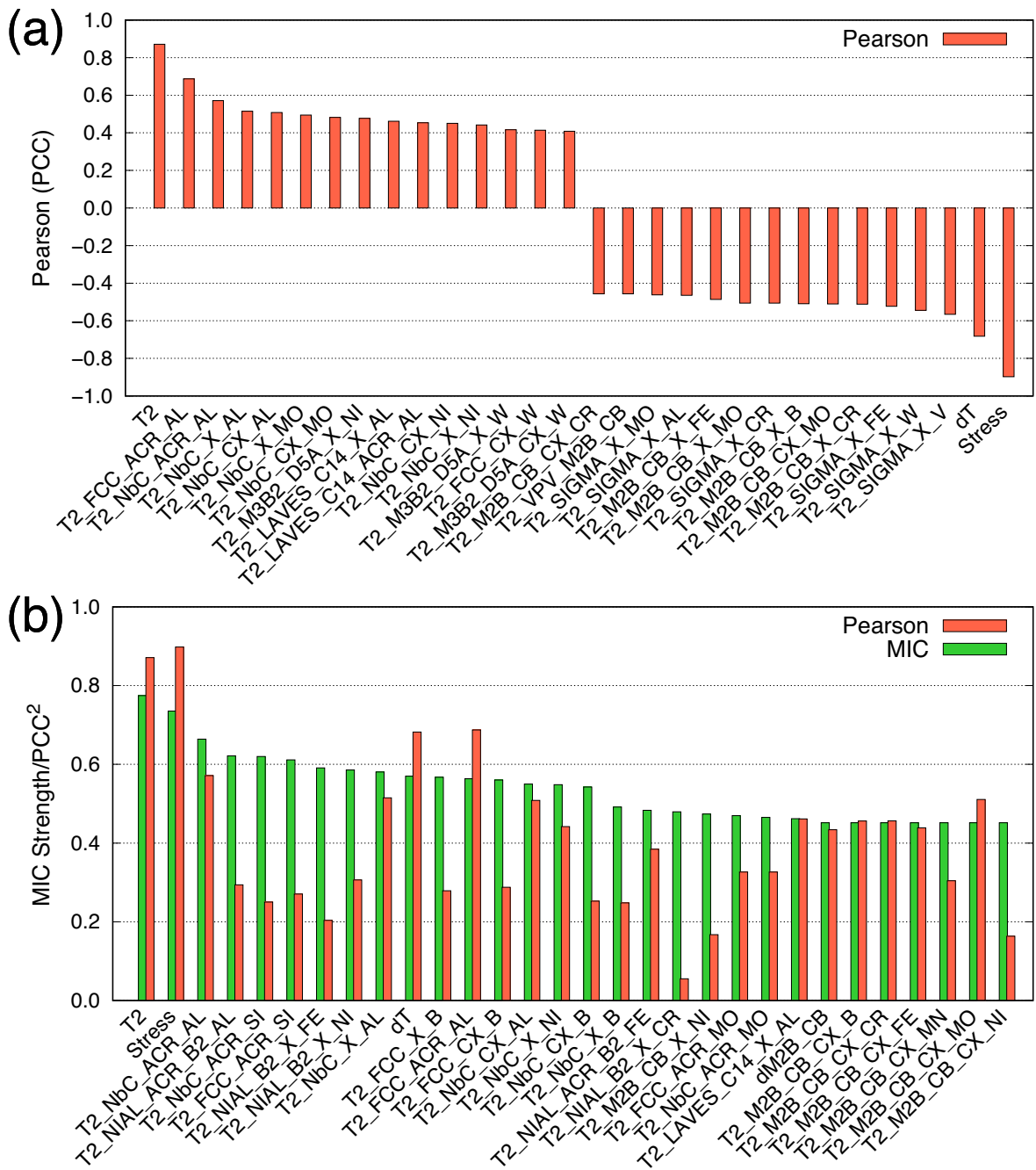

Figure 4 Correlation analysis between all the alloy features (simple/superficial and microstructure-related synthetic/scientific features generated from high-throughput Thermo-Calc computation) and LMP of AFA alloys from (a) Pearson correlation represented as PCC, and (b) from MIC. The MIC result has been compared with PCC$^2$. Top 15 and bottom 15 ranking features out of 466 features are shown in (a), and only top 30 ranking features from MIC analysis and corresponding features from Pearson correlation analysis are shown in (b), respectively.

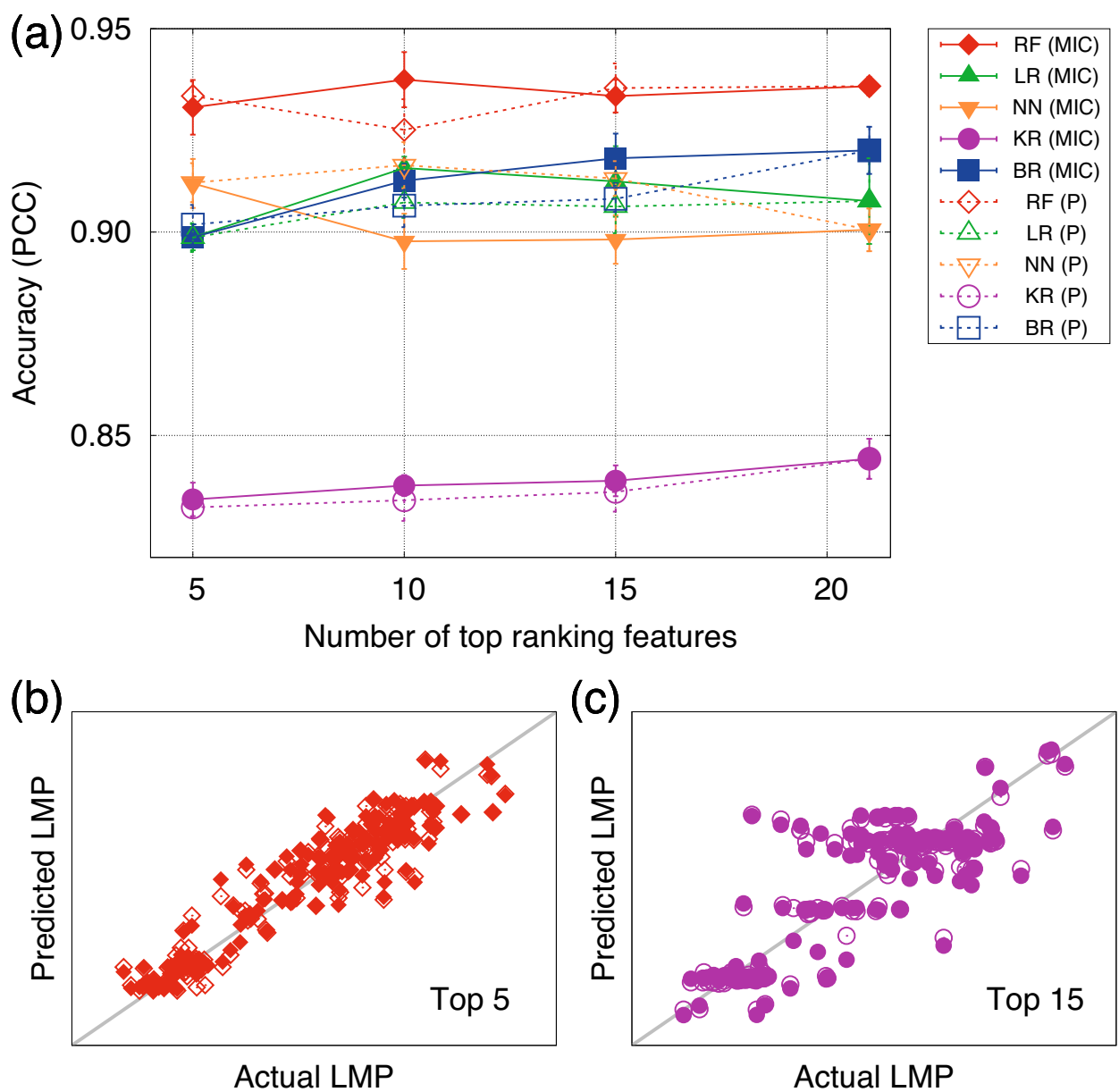

Figure 5 Training results of machine learning models with simple/superficial features, i.e., elemental compositions, stress, solutionizing temperature, and the ratio between Nb and C: (a) accuracy, represented as PCC of five different machine learning models (RF: random forest, LR: linear regression, NN: nearest neighbor, KR: kernel ridge regression, BR: Bayesian ridge regression) as a function of number of top ranking features, (b) actual vs predicted LMP with random forest with top 5 features, (c) actual vs predicted LMP with kernel ridge regression with top 15 features.

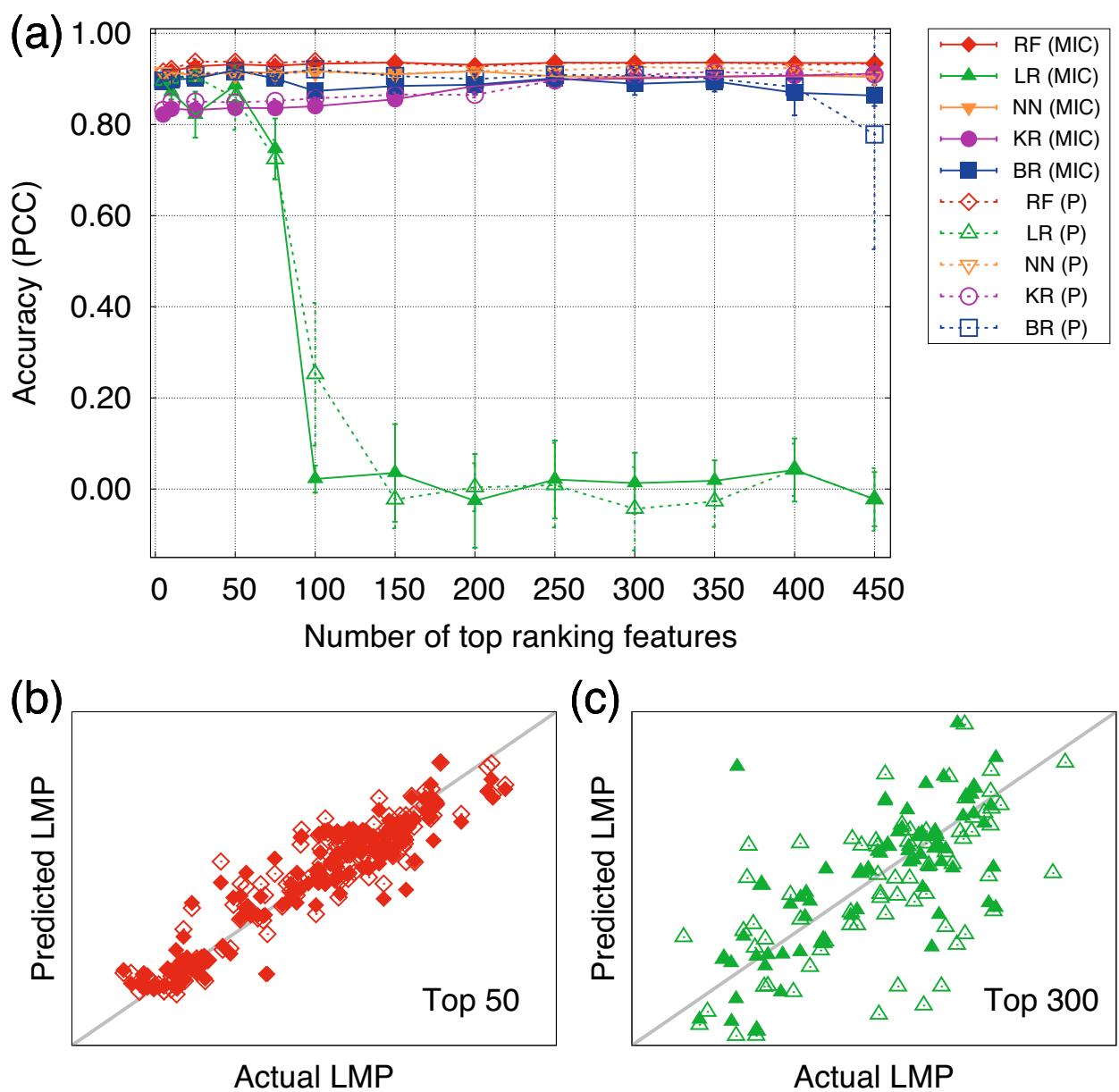

Figure 6 Traninng results of machine learning models with synthetic/scientific features populated from high-throughput computational thermodynamic approach: (a) accuracy, represented as PCC of five different machine learning models (RF: random forest, LR: linear regression, NN: nearest neighbor, KR: kernel ridge regression, BR: Bayesian ridge regression) as a function of number of top ranking features, (b) actual vs predicted LMP with random forest with top 50 features, (c) actual vs predicted LMP with linear regression with top 300 features.

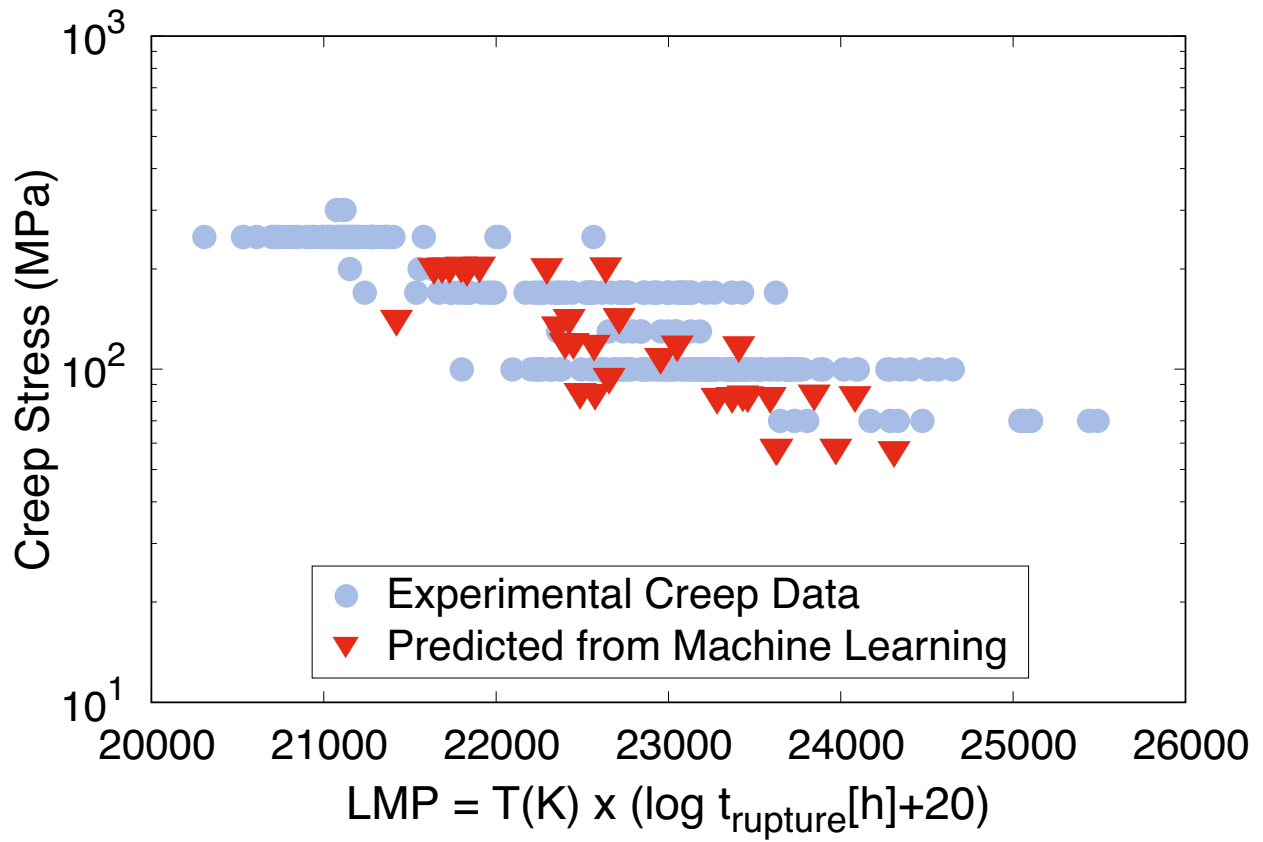

Figure 7 Experimentally measured LMPs of AFA alloys used for the training of machine learning models and predicted LMPs of hypothetical AFA alloys from trained models.